\documentclass[epj]{svjour}
\usepackage{graphics,graphicx}
\begin{document}
\title{Equilibrium and nonequilibrium properties of systems with long-range interactions}
\author{Stefano Ruffo\inst{1,2} 
}                     
%
%
\institute{Dipartimento di Energetica ``S. Stecco" and CSDC, Universit\`a di Firenze 
via s. Marta 3, I-50139 FIRENZE, Italy \and INFN, Sezione di Firenze}
\date{Received: date / Revised version: date}
%
\abstract{
We briefly review some equilibrium and nonequilibrium properties of systems with long-range
interactions. Such systems, which are characterized by a potential that weakly decays at large
distances, have striking properties at equilibrium, like negative specific heat in the 
microcanonical ensemble, temperature jumps at first order phase transitions, broken ergodicity.
Here, we mainly restrict our analysis to mean-field models, where particles globally
interact with the same strength.
We show that relaxation to equilibrium proceeds through quasi-stationary states whose duration
increases with system size. We propose a theoretical explanation, based on Lynden-Bell's entropy, 
of this intriguing relaxation process. This allows to address problems related
to nonequilibrium using an extension of standard equilibrium statistical mechanics.
We discuss in some detail the example of the dynamics of the free electron laser, where
the existence and features of quasi-stationary states is likely to be tested experimentally
in the future. We conclude with some perspectives to study open problems and to find
applications of these ideas to dipolar media.
} 
\maketitle
\section{Introduction}
\label{intro}
For systems with long-range interaction the pair potential decays at large
distances with a weak power law $V(r) \sim r^{-s}$, with $s \leq d$,
the space dimension \cite{Leshouches}. Examples are: gravity, whose statistical mechanics is made
more complex by the unremovable {\it singularity} of the potential at the origin; Coulomb interactions,
with the phenomenon of {\it charge screening} which allows the treatment of globally neutral
systems; dipolar media, that display the well known feature of {\it shape dependence};
vortices interacting with logarithmic potential in $d=2$, for which Onsager first
discussed microcanonical features, like the presence of {\it negative temperatures}, at
the first Statphys meeting in Florence \cite{Onsager}.

\begin{figure}
\resizebox{0.5\columnwidth}{!}{%
  \includegraphics{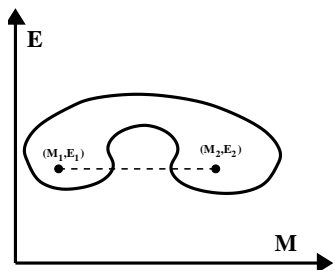}
}
\caption{The set of accessible macrostates in the $(E,M)$ space can have a non-convex
shape for systems with long-range interactions, such that if $(E_1,M_1)$ and $(E_2,M_2)$
can be realized macroscopically, this is not necessarily true for all the states joining
these two along the straight dashed line.}
\label{con}
\end{figure}

Long-range interactions can be made extensive, but are intrinsically non-additive.
Let us consider the simplest model of ferromagnetic systems, the Curie-Weiss
mean-field Hamiltonian
\begin{equation}
H=-\frac{J}{2N} \sum_{i,j} \sigma_i \sigma_j
\label{Curie}
\end{equation}
in which the spins $\sigma_i=\pm 1$ are globally coupled with strength $J>0$.
The energy scales with system size, $H \sim N$, and an intensive energy density 
${\cal E}=\lim_{N \to \infty} H/N$ can be defined in the thermodynamic limit, 
but, due to the presence of $N^2$
links, the sum of the energies of two subsystems $I$ and $II$ is never giving the
total energy $E_{I+II} \neq E_I+E_{II}$. 
Besides that, for mean-field models of this kind the set of accessible macrostates in
the space of intensive parameters $E$ and $M=\sum_i \sigma_i$, can be {\it non convex} 
(see Fig.~\ref{con}). These two mathematical properties, that can be present only at finite
$N$ for short-range interactions ($s > d$), have important physical consequences.
The specific heat \cite{Lynden}, and other quantities related to the curvature of the entropy,
like susceptibility \cite{Touchette}, can become negative in certain energy ranges; jumps in temperature
can be realized at first-order microcanonical transitions \cite{Barre-01}; 
broken ergodicity can be present both for finite $N$ and in the thermodynamic limit
\cite{Mukamel-05} (see Sect.~\ref{broken}). All this is inscribed inside the general framework of {\it ensemble
inequivalence} \cite{Ensemble,Ispolatov}. We will briefly illustrate these features for
a mean-field XY model with two-spin and four-spin interactions in Sect.~\ref{XY}.  

All the above is for equilibrium properties (i.e. maximal entropy states), but systems with 
long-range interactions also show a very slow approach to equilibrium. For systems 
with short-range interactions it has been definitely assessed that, appropriately selecting 
the initial state, the subsequent relaxation to equilibrium takes place on a finite time 
in the thermodynamic limit \cite{Lichtenberg}. On the contrary, since
the seminal paper of Lynden-Bell \cite{Lynden}, it has been proposed that systems with
long-range interactions can display a two-step relaxation. In a first
stage, the system relaxes rapidly (``violently", according to Lynden-Bell) to a {\it quasi-stationary} 
state whose lifetime increases with system size: two types of dependencies have been proposed,
either power-law $N^\delta$ \cite{Yama-04} or logarithmic $\ln N$ \cite{Mukamel-05},
depending on some detailed property of the initial state. In a second stage, the system begins
a slow approach to equilibrium that may be either direct or proceed through successive
relaxations to different quasi-stationary states. The fact that the lifetime of quasi-stationary
states diverges with system size allows to obtain a separation of the two time scales.
This scenario, originally proposed for gravity, has been extended to the two-dimensional 
Euler equation, the original Onsager's system, by Chavanis \cite{Chava-96}.
A theoretical proposal has been advanced by Lynden-Bell in order to interpret the initial relaxation 
to a quasi-stationary state: that the system tries to maximize an entropy which takes into account
additional ``constraints" appearing in the thermodynamic limit, the most relevant of those
being the normalization of the one-body distribution function \cite{Lynden}. Although the initial
numerical tests of this theory gave some hope of success \cite{Hohl}, its predictions were
subsequently disproved for several systems, even for the simple one-dimensional self-gravitating
system \cite{Sakagami} and the theory fell into some discredit. We will show in Sect.~\ref{QS} results 
for the so-called Hamiltonian Mean Field (HMF) model \cite{Antoni} (a mean-field XY model) for which 
Lynden-Bell theory has a 
straightforward application and leads to predictions that are in reasonable agreement 
with numerical simulations
\cite{Chava-06}. A recent domain of application of Lynden-Bell's ideas is to the 
free electron laser \cite{FEL}, where this approach allows to predict the features of the
intensity saturation of the laser, simply knowing the initial conditions, without explicitly
solving the equations of motion, as was tipycally done before: in a sense, this constitutes 
the statistical mechanics of the free electron laser. This application is likely to produce
in the future experimental tests of the features of quasi-stationary states, as we discuss
in Sect.~\ref{Application}. 

Furthermore, the macrostates that maximize Lynden-Bell's entropy for a given energy depend 
on the initial magnetization of the HMF model. The system undergoes a nonequilibrium phase transition that
can be of the second or the first order, and hence a {\it nonequilibrium tricritical point} is present
(see Sect.~\ref{TP}). It is interesting that the concepts of equilibrium phase transition can
be extended to states that are not in equilibrium. It means that one can assume {\it mixing} limited
to the phase-space visited by the trajectories on a time scale that is small with respect 
to the relaxation time to equilibrium.
 
Sect.~\ref{Conclusions} is devoted to some final remarks and perspectives of applications
to wave-particle systems and to dipolar media.

\section{Phase diagram of a mean-field XY model}
\label{XY}

\begin{figure}
\resizebox{0.75\columnwidth}{!}{%
  \includegraphics{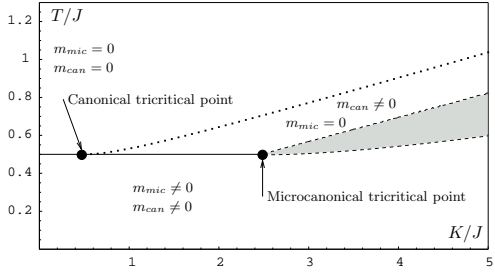}
}
\caption{Phase diagram of Hamiltonian (\ref{XYeq}). The canonical second order transition
line (full horizontal line starting at $T/J=1/2$) becomes first order (dotted line) at the canonical 
tricritical point.
The microcanonical second order transition line coincides with the canonical one up to
$K/J=1/2$ but it extends further towards the microcanonical tricritical point, located at $K/J=5/2$.
At this latter point, the transition line bifurcates in two first order microcanonical lines (dashed),
corresponding to a temperature jump. There are no microcanonical macrostates for parameter values 
within the shaded region.}
\label{dia}
\end{figure}

\begin{figure}
\resizebox{0.75\columnwidth}{!}{%
  \includegraphics{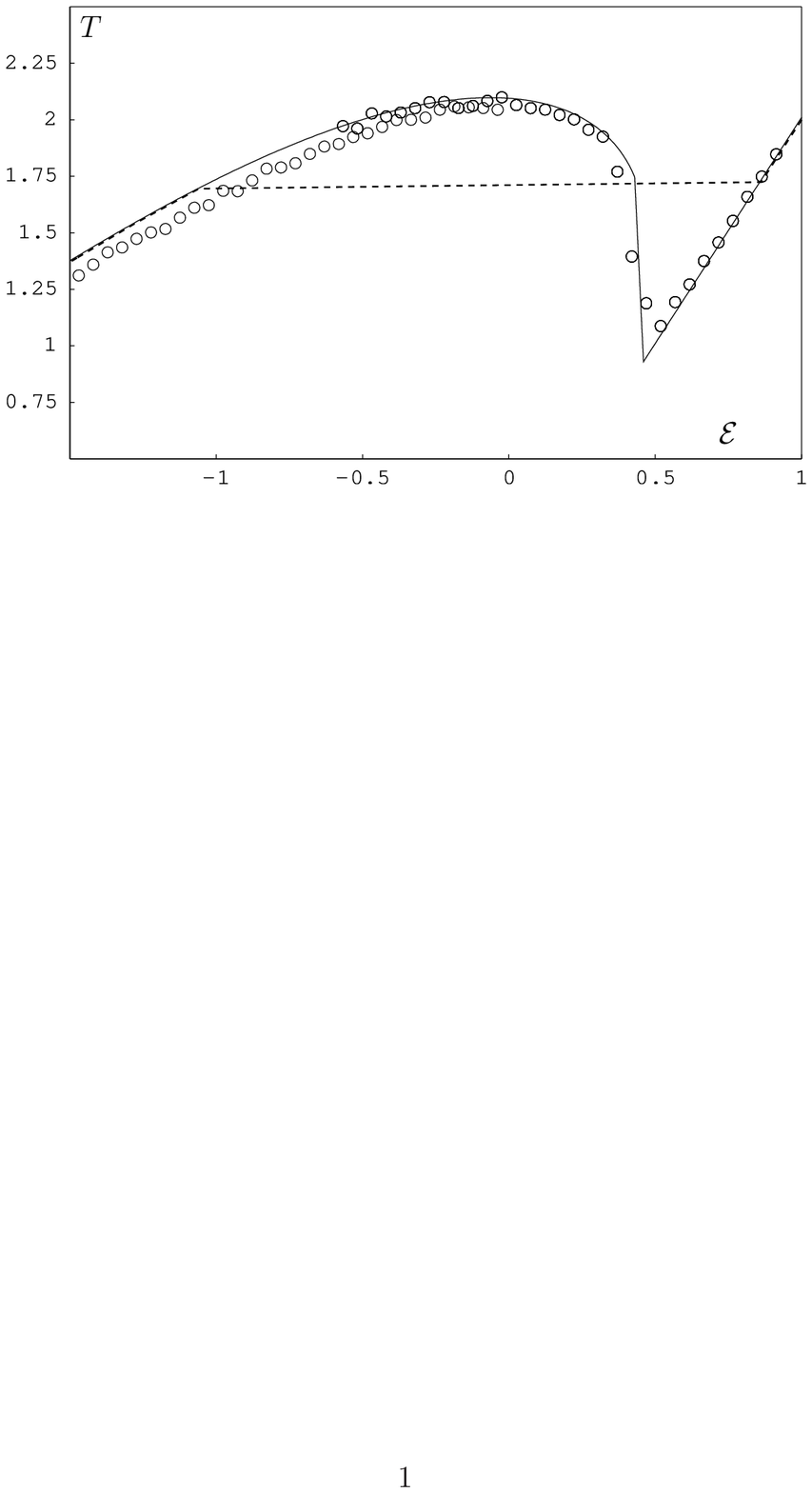}
}
\caption{Caloric curve for $K/J=10$. The full line is the theoretical prediction in the
microcanonical ensemble. The dashed line represents the first order phase transition in
the canonical ensemble. The points are obtained from a molecular dynamics simulation of 
Hamiltonian (\ref{XY}) with $N=100$.}
\label{cal}
\end{figure}

Let us consider the following Hamiltonian
\begin{equation}
H_{XY}=\sum_{i=1}^{N}\frac{p_i^2}{2}
-\frac{J}{2N}(\sum_{i=1}^{N}\vec s_i)^2
-\frac{K}{4N^3}\left[(\sum_{i=1}^{N}\vec s_i)^2\right]^2,
\label{XYeq}
\end{equation}
which can be thought as representing a system of $N$ spins 
$\vec s_i=(\cos \theta_i, \sin \theta _i)$ with all-to-all
two-spin, $J$, and four-spin, $K$, interactions. A kinetic energy term 
is added, considering $p_i$ as canonically conjugate to $\theta_i$.
Because of this addition, Hamiltonian (\ref{XYeq}) can also
represent a system of unit mass particles moving
on a circle without collisions, interacting only through a mean-field type potential.
This model can be solved in both the canonical and the microcanonical
ensemble \cite{Debuyl}. The resulting phase diagram is shown in
Fig.~\ref{dia}. For both ensembles a tricritical point is present,
but its location is different in the two ensembles. The behavior of the 
order parameter $m=\lim_{N \to \infty}|\sum_i \vec s_i|/N$  in the two ensembles 
is also shown, in order to highlight the striking difference 
in the predictions. The so-called {\it caloric curve} (kinetic temperature vs. 
energy density ${\cal E}$) is reported in Fig.~\ref{cal}. 
The microcanonical ensemble (full line)
predicts a region of negative specific heat, where kinetic temperature $T$ decreases
as the energy density is increased. Moreover, a temperature jump is present at 
the transition energy. 

\section{Quasi-stationary states}
\label{QS}

\begin{figure}
\resizebox{0.75\columnwidth}{!}{%
  \includegraphics{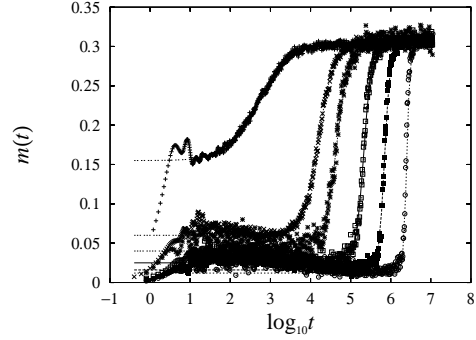}
}
\caption{Magnetization $m$ vs. time $t$ for the HMF model with energy ${\cal E}=0.69$, 
vanishing total momentum
and ``water-bag" initial condition with $m_0=0$ ($\Delta \theta = \pi$). 
The value of $N$ increases from left to right:
$N=10^2,10^3,2 \times 10^3, 5 \times 10^3, 10^4, 2 \times 10^4$.} 
\label{longtime}
\end{figure}

Hamiltonian (\ref{XYeq}) reduces, for $K=0$ and adding a constant to shift the
energy of the ground state to zero, to that of the HMF model \cite{Antoni}
\begin{equation}
H_{HMF}=\sum_{i=1}^{N}\frac{p_i^2}{2}
+\frac{1}{2N}\sum_{i,j=1}^{N} (1-\cos (\theta_i-\theta_j)),
\label{HMF}
\end{equation}
whose equilibrium properties are standard: the system undergoes a mean-field
second order phase transition in both the microcanonical and the canonical
ensemble at the energy ${\cal E}=3/4$, corresponding to the temperature $T=1/2$
(see \cite{Vatteville} for a comprehensive recent review of the model).
In order to study nonequilibrium properties, it has been customary to prepare
an initial state of the ``water-bag" type where all the particles are uniformly
distributed in a rectangular domain of width $2 \Delta \theta$ and height 
$2 \Delta p$ centered around the origin in the single-particle phase space
$(\theta,p)$. Once the size of the domain is given, energy and magnetization
are uniquely determined: $m_{0} = {\sin \Delta\theta }/{\Delta\theta}$,
${\cal E}= (\Delta p)^{2}/6 + (1-m_{0}^{2})/{2}$. During microcanonical time evolution,
energy and total momentum $\sum_i p_i$ remain constant, but magnetization varies, and 
one expects that it reaches the equilibrium value compatible with the given energy. 
This is not what happens, as
shown in Fig.~\ref{longtime}. One observes an initial ``violent" relaxation 
to a plateau value (closer and closer to $m=0$ as the number of particles
is increased), corresponding to the quasi-stationary state, followed by a second 
relaxation to equilibrium, which takes place on longer and longer times as $N$ increases. 
The lifetime of the quasi-stationary state has been fitted with a power-law
$N^{1.7}$ in Ref.~\cite{Yama-04}. 

\begin{figure}
\resizebox{0.75\columnwidth}{!}{%
  \includegraphics[angle=0]{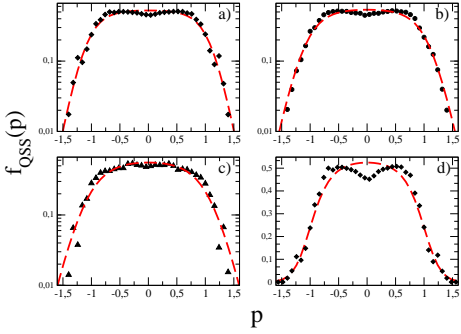}
}
\caption{(color online) Comparison of the theoretical predictions for the quasi-stationary
momentum distribution (dashed line) with the numerical simulations (points) 
performed using Hamiltonian (\ref{HMF}) for ${\cal E}=0.69$ 
and $m_0=0.3$ (a), $m_0=0.5$ (b), $m_0=0.7$ (c) . The case $m_0=0.3$ is also shown in
lin-lin scale in panel (d). For all these cases $m[\bar{f}_{QSS}]=0$ (homogeneous 
quasi-stationary states).}
\label{distr}
\end{figure}

\begin{figure}
\resizebox{0.75\columnwidth}{!}{%
  \includegraphics{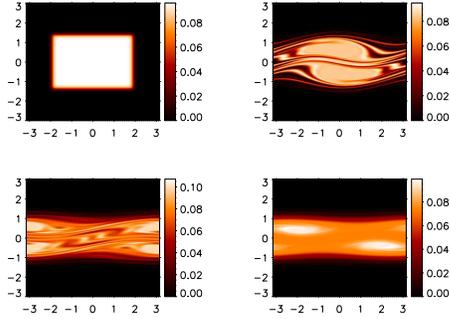}
}
\caption{(color online) Time evolution of the single-particle distribution function according to
the Vlasov equation (\ref{vlasoveq}) for a rectangular ``water-bag" initial state with 
${\cal E}=0.69$ and $m_0=0.5$.}
\label{fluid}
\end{figure}

How does one explain such a behaviour? It is
crucial to realize that the mean-field dynamics (\ref{HMF}) is well represented 
in the large $N$ limit
by the following Vlasov equation \cite{BraunHepp}
\begin{equation}
\frac{\partial f}{\partial t} + p\frac{\partial f}{\partial \theta} -
\frac{d V}{d \theta} \frac{\partial f}{\partial p}=0,
\label{vlasoveq}
\end{equation}
where $f(\theta,p,t)$ is the single-particle distribution function and 
the potential $V(\theta)$ is given by
\begin{eqnarray*}
V(\theta)[f] &=& 1- m_x[f] \cos(\theta) - m_y[f] \sin(\theta), \\
m_x[f] &=& \int   f(\theta,p,t) \, \cos{\theta}  {\mathrm d}\theta
{\mathrm d}p, \\
m_y[f] &=& \int   f(\theta,p,t) \, \sin{\theta}{\mathrm d}\theta
{\mathrm d}p.
\end{eqnarray*}
Besides energy, Vlasov equation also conserves the norm of the distribution
function $\int f(\theta,p,t) d \theta dp = 1$, which for the ``water-bag" initial condition
we have considered implies that the distribution function remains two-level
$(0,f_0=1/(4 \Delta \theta \Delta p))$ at all times. To take this into account,
Lynden-Bell has proposed \cite{Lynden} that the system adapts itself on macrostates that
maximize the following ``fermionic" entropy  
\begin{equation}
\label{LBentropy}
s_{LB}(\bar{f})=-\int \!\! dp d\theta \,
\left[\frac{\bar{f}}{f_0} \ln \frac{\bar{f}}{f_0}
+\left(1-\frac{\bar{f}}{f_0}\right)\ln
\left(1-\frac{\bar{f}}{f_0}\right)\right],
\end{equation}
where ${\bar{f}}$ is the coarse-grained distribution function. The Pauli principle that
is implicit in this approach refers to the fact that a ``fluid element" in the $(\theta,p)$
$\mu$-space cannot occupy a cell which is already occupied by another fluid element, just because
the distribution must remain two-level in the course of time.
It turns out that the maximization of $s_{LB}(\bar{f})$ at fixed energy, momentum and norm can be
explicitly performed for the HMF model, giving the following solution
\begin{equation}
\label{barf}
\bar{f}_{QSS}(\theta,p)=
\frac{f_0}{e^{\beta (p^2/2 -m[\bar{f}_{QSS}]\cos\theta)+\lambda p+\alpha}+1},
\end{equation}
where $\beta$, $\lambda$ and $\alpha$ are Lagrange multipliers corresponding to
the conservation of energy, momentum and norm. As it is clear from equation (\ref{barf}),
the quasi-stationary distribution has to be determined self-consistently, since 
it depends on the distribution itself through the magnetization $m[\bar{f}_{QSS}]$.
We have chosen the subscript QSS to mean
quasi-stationary-state, because we propose \cite{Chava-06} that the states that maximize Lynden-Bell
entropy are indeed the quasi-stationary states observed, e.g., in the numerical
simulation reported in Fig.~\ref{longtime}. The magnetization in the QSS,
$m[\bar{f}_{QSS}]=m_{QSS}$, and the values of the Lagrange multipliers 
are obtained by solving numerically a set of implicit equations \cite{Chava-06}, once
the energy ${\cal E}$ and the initial magnetization $m_0$ are given (or alternatively
$(\Delta \theta, \Delta p)$). Since we look for solutions whose 
total momentum vanishes, $\lambda=0$. In Fig.~\ref{distr} we show the comparison of
the predictions of the theory concerning momentum distributions with the numerical simulations  
performed integrating the equations of motion given by Hamiltonian (\ref{HMF}) with $N=10000$. 
The agreement is quite good, if one takes into account that the predictions have no 
adjustable parameter and are strictly determined by the choice of the initial condition. 
However, expecially the plot in lin-lin scale (panel (d) in Fig.~\ref{distr}) reveals 
that our theory is unable to reproduce the double bump obtained in numerical simulations.
The presence of this bumpy feature is confirmed by simulating the Vlasov equation
(\ref{vlasoveq}) (see Fig.~\ref{fluid}) \cite{Califano}. The double bump in momentum distribution
is the result of the presence of two sliding resonances (the two vortices in the lower
right panel), for which an explanation that takes into account the specific dynamical
properties of the model is necessary \cite{Chava-06,Califano}.

Quasi-stationary states have been demonstrated to be robust to both the application
of and external bath \cite{Baldovin} and to the addition of a small nearest neighbor
interaction \cite{Giansanti}.
The Vlasov equation is exact in the $N \to \infty$ limit, and people have tried to
address the question of finite $N$ corrections. In this respects the most interesting
progress is registered in papers by Bouchet and Dauxois \cite{BouchetDauxois} and
Chavanis \cite{Chava-Vlasov}.
For the sake of commpleteness, it should be mentioned that a different approach, based 
on Tsallis statistics, has been proposed to describe quasi-stationary states \cite{Rapisarda}. 
Recently, these authors have concentrated their attention on the behaviour of correlations
in the single particle dynamics, that can give rise to non Gaussian distributions for 
sums of variables (generalized central limit theorems).

\section{Nonequilibrium tricritical point}
\label{TP}

\begin{figure}
\resizebox{0.75\columnwidth}{!}{%
  \includegraphics[height=3truecm,width=5truecm]{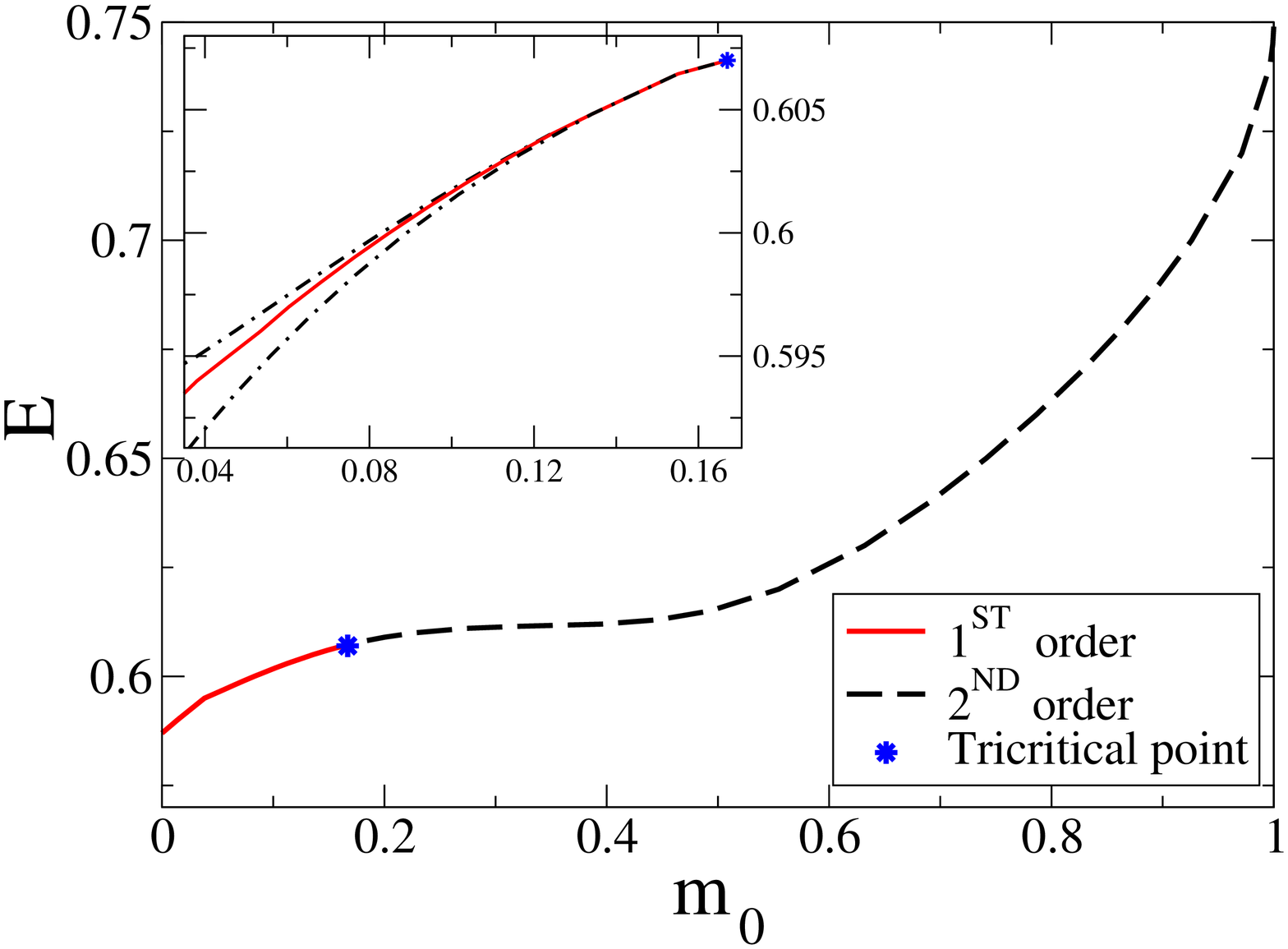}
}
\caption{(color online) Theoretical phase diagram on the control parameter plane $(m_{0},{\cal E})$: 
second order phase transition line (dashed); first order phase transition line (full); 
tricritical point (full dot). Inset: magnification of the first order 
phase transition region and limits of the metastability region (dash-dotted).}
\label{tric}
\end{figure}

The maximization of Lynden-Bell entropy reserves another surprise. Since this variational
problem introduces another control parameter besides energy, the magnetization $m_0$ of 
the initial state, one obtains a phase transition at an energy that depends on $m_0$. The
transition energy coincides with that given by Boltzmann entropy (to which Lynden-Bell's
entropy reduces in the diluted limit $\bar{f} \ll f_0$) only for $m_0=1$. Not only, below
$m_0=0.61$ the transition becomes first order. The full phase diagram is plotted in
Fig.~\ref{tric}. A {\it nonequilibrium tricritical point} is present in the phase 
diagram. The specific heat is negative along all the transition line and a temperature 
jump appears when the transition change to first order.

\begin{figure}
\resizebox{0.75\columnwidth}{!}{%
  \includegraphics[height=4truecm,width=6truecm]{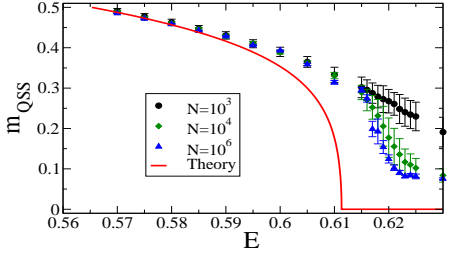}
}
\caption{(color online) $m_{QSS}$ as a function of ${\cal E} $ for 
$m_{0}=0.30$ where the phase transition is second order. The full line is the
theoretical prediction and the points are simulations of the HMF model for which
the number of initial realizations is $10^5$ ($N=10^3$), $10^4$ ($N=10^4$), $10^2$ ($N=10^6$) 
and the averaging time $20 < t \leq 100$.}
\label{trans_a}
\end{figure}

\begin{figure}
\resizebox{0.75\columnwidth}{!}{%
  \includegraphics[width=6truecm,height=4truecm]{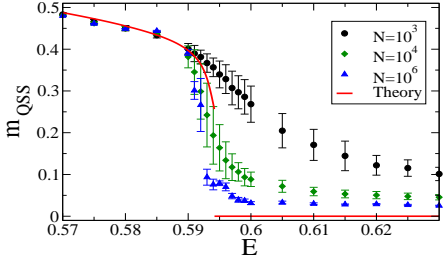}
}
\caption{(color online) $m_{QSS}$ as a function of ${\cal E} $ for 
$m_{0}=0.05$ where the phase transition is first order. The results are plotted as in 
Fig.~\ref{trans_a}.}
\label{trans_b}
\end{figure}

Numerical simulations confirm this extremely interesting theoretical finding 
\cite{Antoniazzi}. For instance, we plot in Figs.~\ref{trans_a},\ref{trans_b} the order parameter
$m_{QSS}$ as a function of energy density at fixed $m_0$, finding signature of a 
second order (Fig.~\ref{trans_a}) or first order (Fig.~\ref{trans_b}) phase transition.
As previously explained, these are ``short-time" phase transitions, at variance with
the ``long-time" phase transitions of equilibrium statistical mechanics. Hence, in order 
to reveal them numerically, one has to average over a short initial time and take many
initial instances. 

\section{Broken ergodicity}
\label{broken}

\begin{figure}
\resizebox{0.75\columnwidth}{!}{%
  \includegraphics[angle=0]{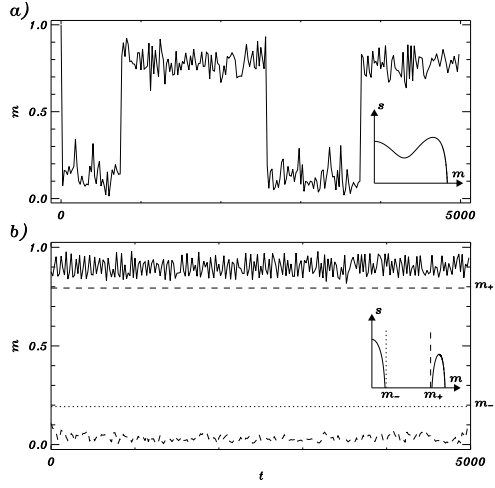}
}
\caption{Time evolution of the magnetization $m$  of model (\ref{XYeq}) for:
a) ${\cal E}=0.1$, $J=-1$ and $K=8$; b)  ${\cal E}=0.1077$, $J=-1$ and $K=3$,
obtained integrating the equations of motion derived from Hamiltonian
(\ref{XYeq}) with $N=20$.
In panel a) magnetization switches between the two most probable values
(see the inset for the dependence of entropy on $m$).
In panel b), two different initial conditions are plotted
simultaneously, corresponding to two different values of the initial
magnetization, $m_0=0.1$ and $m_0=0.98$. The inset shows the entropy, which now
vanishes in the interval $[m_-,m_+]$. Here $m_-\simeq 0.192$ and $m_+\simeq 0.794$, which
are shown to bound from above (dotted line) and from below (dashed line) the
magnetization of the two initial conditions.}
\label{break}
\end{figure}

After this excursion in nonequilibrium, let us come back to equilibrium properties.
The XY model (\ref{XYeq}), for some specific choice of antiferromagnetic $J<0$ and
ferromagnetic $K>0$ couplings has energy density values where regions with different magnetization
$m$ are disconnected if energy is fixed (as shown in Fig.~\ref{con}) \cite{broken}. An example is 
given in Fig.~\ref{break},
where we show two different situations. In panel a) the ``standard" ergodicity breaking associated with
a phase transition is revealead by successive magnetization switches between local entropy
maxima: the number of particles is small ($N=20$) such that the entropy barrier
(see inset) is not insurmontable. In panel b) we display the new type of {\it microcanonical
broken ergodicity} first discussed in Refs.~\cite{Mukamel-05,Borgonovi}: now magnetization cannot
switch because there is a gap $[m_-,m_+]$ with no macrostates.

\section{Application to the free electron laser}
\label{Application}

Free-Electron Lasers (FELs) are coherent and tunable
radiation sources, which differ from conventional lasers because they use a 
relativistic electron beam as their lasing medium, hence the term ``free-electron".
The physical mechanism responsible for light emission and
amplification is the interaction  between the relativistic electron
beam, a magnetostatic periodic field generated in the 
undulator and an optical wave co-propagating with the 
electrons. Due to the presence of the magnetic  field, electrons are forced to follow 
oscillating (tipycally sinusoidal) trajectories and emit synchrotron radiation. This  
{\it spontaneous emission} is then amplified along the undulator until the laser effect is reached.
Among the different schemes, single-pass high-gain FELs are currently attracting a growing
interest. Basic features of the system dynamics are successfully captured by a simple  
Hamiltonian model introduced by Colson and Bonifacio \cite{Bonifacio}
\begin{eqnarray}
\frac{d \theta_j}{d \bar{z}} &=& p_j \nonumber\\
\frac{d p_j}{d \bar{z}} &=&-\mathbf{A}e^{i\theta_j}-\mathbf{A}^{\ast}e^{-i\theta_j} \nonumber\\
\frac{d \mathbf{A}}{d \bar{z}} &=& i\delta\mathbf{A}+\frac{1}{N}\sum_j e^{-i\theta_j},
\label{FELeq}
\end{eqnarray}
where $\bar{z}=2k_w \rho z \gamma_r^2/\langle \gamma \rangle_0^2$ is the
rescaled longitudinal coordinate, which plays the role of time.
Here, $\rho=\gamma_r^{-1}(\frac{a_{w} \omega_p}{4 c k_w})^{2/3}$ is the so-called
Pierce parameter, $\langle \gamma \rangle_0$ the mean energy of the electrons 
at the undulator's entrance, $k_w=2 \pi / \lambda_w$ the wave 
number of the undulator, $\omega_p=(4 \pi e^2n/m)^{1/2}$ the plasma frequency, $c$
the speed of light, $n$ the total electron number density,
$e$ and $m$ respectively the charge and mass of one electron. Furthermore, 
$a_w=eB_w/(k_w m c^2)$, where $B_w$ is the
rms peak undulator field. Here $\gamma_r=\left( \lambda_w  
(1+a_w^2)/2 \lambda \right)^{1/2}$ is the resonant energy,  $\lambda_w$ and $\lambda$
being respectively the period of the undulator and the wavelenght of
the radiation field. Introducing  the wavenumber $k$ of the FEL
radiation, the two canonically conjugated variables are
($\theta$,$p$), defined as $\theta=(k+k_w)z-2\delta\rho k_w z \gamma_r^2/\langle \gamma
\rangle_0^2$ and $p=(\gamma-\langle \gamma \rangle_0)/(\rho \langle \gamma \rangle_0)$. 
$\theta$ corresponds to the phase of the electrons with respect to the ponderomotive wave. 
The complex amplitude ${\bf A}=A_x + iA_y $ represents the
scaled field, transversal to $z$.  Finally, the detuning
parameter is given by $\delta=(\langle \gamma
\rangle_0^2-\gamma_r^2)/(2\rho\gamma_r^2)$, and measures the
average relative deviation from the resonance condition. This succinct, but detailed, description of the
model should transmit the idea that, by adjusting model parameters, one can really simulate
realistic experimental situations.

It can be easily checked that model (\ref{FELeq}) can be derived from the Hamiltonian
\begin{equation}
H_{FEL}=\sum_{j=1}^N\frac{p_j^2}{2} -N \delta I +2\sqrt{I}\sum_{j=1}^N
\sin(\theta_j-\varphi) 
\label{FELHam},
\end{equation}
where $N$ is the number of electrons and the intensity~$I$ and
the phase~$\varphi$ of the wave are related to $\mathbf{A}=A_x+iA_y=\sqrt{I}\ e^{-i\varphi}$. 
In addition to the energy, the total momentum $P=\sum_j p_j + N \mathbf{A}\mathbf{A}^{\ast}$ is
also a conserved quantity.

It should also be mentioned that, remarkably, this simplified formulation applies
to other physical systems, provided an identification of the
variables involved is performed. As an example, consider the electron beam-plasma instability. 
When a weak electron beam is injected into a thermal plasma,
electrostatic modes at the plasma frequency (Langmuir modes) are
destabilized.  The interaction of the Langmuir waves
and the electrons constituting the beam can be studied in the framework
of a self-consistent Hamiltonian picture \cite{ElskensBook}, formally equivalent to
the one in \cite{Bonifacio}. In a recent paper \cite{andrea}  we have established 
a bridge between these two areas of investigation and exploited the connection to 
derive a reduced Hamiltonian model to characterize the saturated dynamics of the laser.  

\begin{figure}
\resizebox{0.75\columnwidth}{!}{%
  \includegraphics[angle=0]{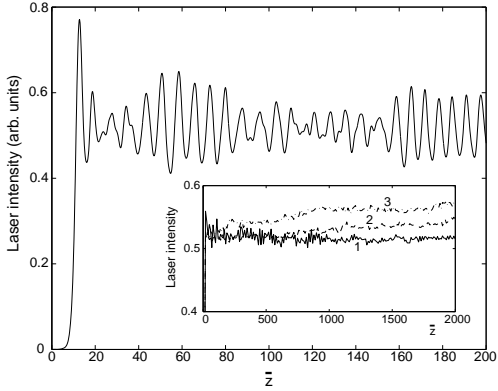}
}
\caption{Typical evolution of the radiation
intensity of a free electron laser using Eqs.~(\ref{FELeq}); the
detuning $\delta$ is set to $0$, the energy per electron $H/N=0.2$ and
$N=10^4$ electrons are simulated. The inset presents averaged
simulations on longer times for different values of $N$:
$5\cdot10^3$ (curve 1), 400 (curve~2) and 100 (curve~3).}
\label{saturation}
\end{figure}

There are many similarities between model (\ref{FELHam}) and the HMF model (\ref{HMF}),
and indeed the canonical free energy of one model can be exactly mapped onto the other \cite{largedev}.
While in the HMF model particles interact directly, in the FEL dynamics electrons 
interact only through the field $\mathbf{A}$, whose dynamics depend in turn on those of
the electrons. 
Therefore, already on this basis one could expect similarities in the behaviours
of the two models. Indeed, similarly to the HMF model, Hamiltonian (\ref{FELHam}) presents
a standard second order phase transition of the mean field type. As for the HMF, FEL
dynamics is well represented in the $N\to\infty$ limit by the following Vlasov
equation
\begin{eqnarray}
\frac{\partial f}{\partial \bar{z}} &=& -p\frac{\partial f}{\partial \theta}
+2(A_x\cos{\theta}-A_y\sin{\theta})\frac{\partial f}{\partial p}, \nonumber\\
\frac{\partial A_x}{\partial \bar{z}}   &=& -\delta
A_y+\frac{1}{2\pi} \int f \cos{\theta}  \, d\theta d p, \nonumber\\
\frac{\partial A_y} {\partial \bar{z}}  &=&  \delta
A_x-\frac{1}{2\pi} \int f \sin \theta \, d\theta dp~.
\end{eqnarray}
Hence, one can similarly expect the existence of quasi-stationary states \cite{FEL}.
A direct simulation of Eqs.~(\ref{FELeq}) is shown in Fig.~\ref{saturation} for
$N=10^4$ electrons.
Initially, the electrons are uniformly distributed in the $\theta$ interval
$[-\pi,\pi]$ and their momentum is also uniformly spread in $[-\Delta p, \Delta p]$
in a ``water-bag" distribution: this initial state has a physical meaning, because
it is very close to the state in which, experimentally, the electrons are injected into
the undulator. The laser intensity $I$ is initially set to a small value, in order to
switch on the instability, whose initial growth is exponential. On a short time,
the intensity reaches a saturation level and performs wide oscillations around it, but
these oscillations dump with time, and a well defined asymptotic intensity is 
reached. This process is the equivalent of Lynden-Bell's ``violent relaxation".
The $N$-dependence of the saturation level is shown in the inset, where three different
values of $N$ are considered. For the smaller value, $N=100$ (curve 3), the saturated
intensity is larger than for higher $N$ values. When $N$ increases the system
remains trapped in the quasi-stationary state for a longer and longer time. The second
relaxation to the larger intensity value corresponds to the relaxation to Boltzmann-Gibbs
equilibrium. This latter relaxation has no experimental relevance for free electron
lasers, because it would take place for enormous undulator lengths.
A similar behavior would be observed for the ``bunching parameter" 
$b = |\sum_n \exp (i \theta_n)|/N$, which is the equivalent of the magnetization $m$
of the HMF  model. 

The distribution $\bar{f}_{QSS}(\theta,p)$ in the quasi-stationary state can be obtained
by maximizing Lynden-Bell entropy (\ref{LBentropy}), keeping energy and momentum $P$
fixed. The result of this maximization procedure is very similar to that of the
HMF
\begin{equation}
\bar{f}_{QSS}(\theta,p)= \frac{f_0}{e^{\beta(p^2/2+2A_{QSS}\sin\theta)+\lambda p+\alpha}+1}~,
\end{equation}
with
\begin{equation}
A_{QSS}=\sqrt{I_{QSS}}=\frac{\beta}{\beta \delta-\lambda}\int dp d\theta\, \sin \theta 
\bar{f}_{QSS}(\theta,p),
\end{equation}
and $\beta$, $\lambda$ and $\alpha$ are the usual Lagrange multipliers, whose value
is determined by the initial condition.

\begin{figure}
\resizebox{0.75\columnwidth}{!}{%
  \includegraphics[angle=0]{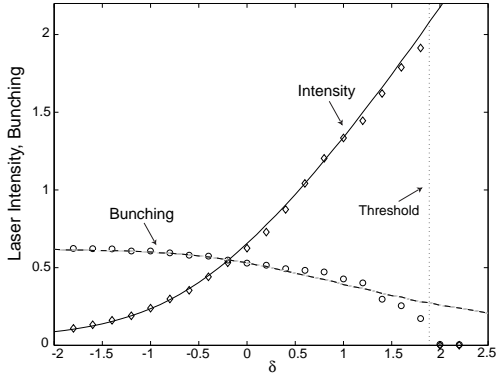}
}
\caption{Laser intensity $I_{QSS}$ and bunching parameter $b_{QSS}$ in the quasi-stationary state 
as a function of detuning $\delta$: theoretical prediction (solid and dashed lines, respectively) 
and simulations (symbols). The dotted vertical line, $\delta=\delta_c\simeq 1.9$, represents 
the transition from the low to the high-gain regime.}
\label{intensity}
\end{figure}

The dependence of both the intensity and the bunching parameter in the quasi-stationary
state on the detuning parameter  $\delta$ for an initially homogeneous state ($\Delta \theta = \pi$)
with zero momentum dispersion $\Delta p=0$ is shown in Fig.~(\ref{intensity}).

\begin{figure}
\resizebox{0.75\columnwidth}{!}{%
  \includegraphics[angle=0]{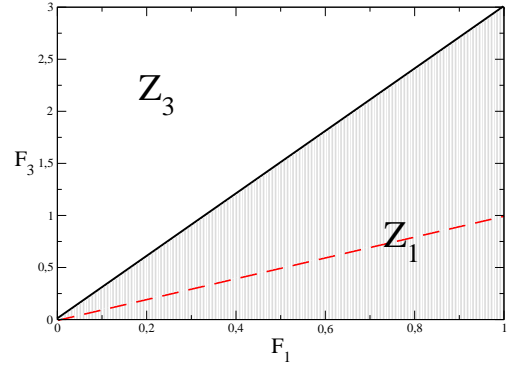}
}
\caption{(color online) Phase diagram in the ($F_1$,$F_3$) parameter space for the two-wave model
(\ref{waves}), corresponding to a FEL with an initial cold homogeneous beam. Lynden-Bell's theory
predicts a transition from a region $Z_1$ (shaded) dominated by the first wave to one, $Z_3$, dominated
by its odd harmonic. Above the dashed line the short time growth rate of $|{\bf A}_3|$ is predicted
to be larger than that of $|{\bf A}_1|$ by a linear theory}
\label{twowaves}
\end{figure}

If FEL's Hamiltonian is equivalent to HMF, one might wonder if situations exist which produce
more complex phase diagrams and dynamical evolutions. Indeed, in the regime where the beam
current and the emittance is small, FEL's dynamics is better described by a model where many planar
waves interact with the beam
\begin{eqnarray}
\frac{{\mathrm d}\theta_j}{{\mathrm d}\bar{z}} &=& p_j \nonumber \\
\frac{{\mathrm d}p_j}{{\mathrm d}\bar{z}} &=&
-\sum_{h}F_h( {\bf A}_he^{i h \theta_j}-{\bf A}_h^{\ast}e^{-i h \theta_j}) \nonumber \\
\frac{{\mathrm d}{\bf A}_h}{{\mathrm d}\bar{z}} &=& \frac{F_h}{N} \sum_j e^{-i h \theta_j},
\label{waves}
\end{eqnarray}
where the complex amplitudes ${\bf A}_h=A_h^x + iA_h^y$ represent the scaled field, transversal
to $\bar{z}$, and the coupling parameters $F_h$ depend on the experimental setup.
We have analysed in some detail \cite{Johal} only the case with two odd harmonics $h=1,3$, performing
the same analysis of the quasi-statonary states as for Eqs.~(\ref{FELeq}). The initial
condition is again a homogeneous almost monocromatic beam, but now there are two ``order parameters",
$|{\bf A}_1|$ and $|{\bf A}_3|$, with the corresponding bunchings 
$b_1=|\sum_j \exp ({-i \theta_j}) /N$ and $b_3=|\sum_j \exp ({-i 3 \theta_j}) /N$.
In region $Z_1$ (shaded) of Fig.~\ref{twowaves} the quasi-stationary state is dominated
by the first wave: hence $|{\bf A}_1| > 0$ and $|{\bf A}_3|=0$. On the contrary, in region
$Z_3$, $|{\bf A}_3| > 0$ and $|{\bf A}_1|=0$. The full line corresponds to a nonequilibrium
phase transition of the first order.

\begin{figure}
\resizebox{0.75\columnwidth}{!}{%
  \includegraphics[angle=0]{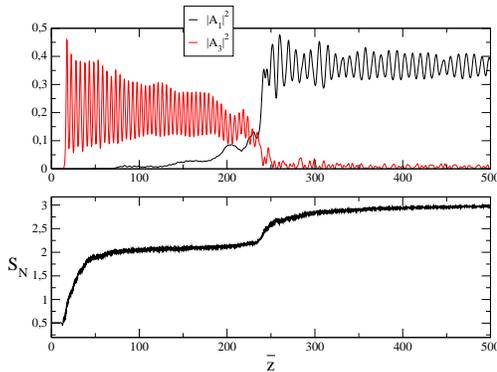}
}
\caption{(color online) Upper panel: Evolution of the intensities $|{\bf A}_1|^2$ and $|{\bf A}_3|^2$ 
showing an initial growth of the first odd harmonic followed by the final relaxation to the
maximal Lynden-Bell's entropy state dominated by the first wave. Lower panel: time
evolution of Lynden-Bell's entropy}
\label{twowaves1}
\end{figure}

However, a linear analysis show that in the region between the dashed line and the transition line 
to $Z_3$ (inside the $Z_1$ region) the growth rate of $|{\bf A}_3|$ is larger than that of $|{\bf A}_1|=0$.
Therefore, at short time the growth of $|{\bf A}_3|$ is expected to be faster and the
initial evolution of the FEL is dominated by the first odd harmonic. However, this is not the
maximal Lynden-Bell's entropy state, and on a later time, the system is finally going to be
dominated by the first wave. This is indeed what happens, as shown in Fig.~\ref{twowaves1}.

\section{Conclusions and perspectives}
\label{Conclusions}

We have discussed both equilibrium and nonequilibrium properties of systems with long
range interactions with reference to mean-field models. Most of these features should
extend to cases where forces weakly decay \cite{Leshouches,largedev}, but this problem
remains to be seriously investigated. In this perspective, the general classification of 
phase transitions obtained in Ref.~\cite{BarreBouchet} could be extremely useful.

The domain of application of these ideas is vast and the perspective to perform key experiments
in the near future is realistic. We have discussed in some detail in Sect.~\ref{Application}
a realistic model of the free electron laser. However, we can foresee applications
to all systems where particles interact with waves in a self-consistent way. As an example,
let us mention collective atomic recoil lasers, where ultracold atoms interact with the
optical modes of the cavity, determining collective effects and first and second order
phase transitions \cite{CARL}.

Another domain where experimental applications are envisageable is that of 
layered spin structures, where dipolar interactions dominate over
Heisenberg exchange \cite{Sievers}. In this case, a
mean-field term of the Hamiltonian is shown to depend on the shape of the sample,
determining the presence or absence of phase transitions \cite{Campa}.

\begin{acknowledgement}
Most of my knowledge of the physics of systems with long-range interactions
has been shaped by J. Barr\'e, F. Bouchet, P.H. Chavanis, T. Dauxois, D. Fanelli
D.H.E. Gross, D. Mukamel, Y.Y. Yamaguchi. I also thank all the coauthors of my papers on this
subject, with whom I have entertained endless discussions. This work is funded by the
PRIN05 grant {\it Dynamics and thermodynamics of systems with long-range interactions}.
\end{acknowledgement}


\begin{thebibliography}{}
%
%

\bibitem{Leshouches}
T. Dauxois, S. Ruffo, E. Arimondo and M. Wilkens (Eds.) \textit{Dynamics
and thermodynamics of systems with long-range interactions} (Springer, Berlin 2001).

\bibitem{Onsager}
L. Onsager, Il Nuovo Cimento (Suppl.) \textbf{6}, (1949) 279.

\bibitem{Lynden} V. A. Antonov, Leningrad Univ. \textbf{7}, (1962) 135 [translation in IAU
Symposium \textbf{113}, (1995) 525];  D. Lynden-Bell, R. Wood, Mon. Not. R. Astr. Soc. 
\textbf{138}, (1968) 495; P. Hertel and W. Thirring, Annals of Physics, \textbf{63}, (1971) 520.

\bibitem{Touchette} A. Campa, S. Ruffo and H. Touchette, Physica A, {\bf 385}, (2007) 233.

\bibitem{Barre-01} J. Barr\'e, D. Mukamel and S. Ruffo, Phys. Rev. Lett.,
\textbf{87}, (2001) 030601.
 
\bibitem{Mukamel-05}  D. Mukamel, S. Ruffo and N. Schreiber, Phys. Rev. Lett. \textbf{95}, (2005) 240604.

\bibitem{Ensemble} M. K. H. Kiessling and J.L. Lebowitz, 
Lett. Math. Phys. \textbf{42}, (1997) 43 and refs. therein; R. S. Ellis, K. Haven and B. Turkington, 
J. Stat.Phys., \textbf{101}, (2000) 999.

\bibitem{Ispolatov} I. Ispolatov and E.G.D. Cohen, Phys. A, \textbf{295}, (2001) 475.

\bibitem{Lichtenberg} J. de Luca, A.J. Lichtenberg and S. Ruffo, Phys. Rev. E, \textbf{60},
(1999) 3781. 

\bibitem{Yama-04} Y.Y. Yamaguchi, J. Barr\'e, F. Bouchet, T. Dauxois,
S. Ruffo, Physica A, \textbf{337}, (2004) 36.

\bibitem{Chava-96} P.H. Chavanis, Ph. D Thesis, ENS Lyon (1996); P.H. Chavanis, J. Sommeria and
R. Robert, Astroph. J., \textbf{471}, (1996) 385.

\bibitem{Hohl} F. Hohl and J.W. Campbell, {\it Collective motion of a one-dimensional self-gravitating
system}, NASA Technical Note TN D-5540, November (1969).

\bibitem{Sakagami} T. Yamashiro, N. Gouda and M. Sakagami, Prog. Theor. Phys., \textbf{88} (1992) 269. 

\bibitem{Antoni} M. Antoni and S. Ruffo, Phys. Rev. E \textbf{52}, (1995) 2361.

\bibitem{FEL} J. Barr{\'e}, T. Dauxois, G. De Ninno, D. Fanelli and S.
Ruffo, Phys. Rev. E, \textbf{69}, (2004) 045501.

\bibitem{Chava-06} P.H. Chavanis, Eur. Phys. J. B \textbf{53}, (2006) 487; 
A. Antoniazzi, D. Fanelli, J. Barr\'e, P.-H. Chavanis, T. Dauxois and S. Ruffo, Phys. Rev. E, 
\textbf{75}, (2007) 011112. 

\bibitem{Debuyl} P. de Buyl, D. Mukamel and S. Ruffo, in \textit{Unsolved Problems of Noise
and Fluctuations}, AIP  Conference Proceedings {\bf 800}, (2005) 533.

\bibitem{Vatteville} P.H. Chavanis, J. Vatteville and F. Bouchet,
Eur. Phys. J. B, \textbf{46}, (2005) 61. 

\bibitem{BraunHepp} W. Braun and K. Hepp, Comm. Math. Phys. \textbf{56}, (1977) 101.

\bibitem{Califano} A. Antoniazzi, F. Califano, D. Fanelli and S. Ruffo, Phys. Rev. Lett., 
\textbf{98}, (2007) 150602.

\bibitem{Baldovin} F. Baldovin and E. Orlandini, Phys. Rev. Lett., \textbf{96}, (2006) 240602;
F. Baldovin and E. Orlandini, Phys. Rev. Lett., \textbf{97}, (2006) 100601.

\bibitem{Giansanti} A. Campa, D. Mukamel, A. Giansanti and S. Ruffo, Phys. A, \textbf{365}, (2006)
120.

\bibitem{BouchetDauxois} F. Bouchet and T. Dauxois, Phys. Rev. E, \textbf{72}, (2005)
045103(R).

\bibitem{Chava-Vlasov} P. H. Chavanis, Phys. A, \textbf{361}, (2006) 55; {\it ibid.}
\textbf{361}, (2006) 81; doi:10.1016/j.physa.2007.10.026, doi:10.1016/j.physa.2007.10.034.


\bibitem{Rapisarda} V. Latora, A. Rapisarda and C. Tsallis, Phys. Rev. E, \textbf{64}, 
(2001) 056134; A. Pluchino, A Rapisarda and C. Tsallis, cond-mat.stat-mech/0706.4021v2.

\bibitem{Antoniazzi}  A. Antoniazzi, D. Fanelli, S. Ruffo and Y.Y. Yamaguchi, 
Phys. Rev. Lett., \textbf{99}, (2007) 040601.

\bibitem{broken} F. Bouchet, T. Dauxois, D. Mukamel and S. Ruffo, {\it Phase space gaps and ergodicity 
breaking in systems with long range interactions}, arXiv:0711.0268.

\bibitem{Borgonovi} F. Borgonovi, G.L. Celardo, M. Maianti and E. Pedersoli, J. Stat. Phys.,
\textbf{116}, (2004) 1435.

\bibitem{Bonifacio} W. B. Colson, Phys. Lett. A,, \textbf{59}, (1976) 187;
R. Bonifacio et al., Opt. Comm., {\bf 50}, (1984), 373.

\bibitem{largedev} J. Barr\'e, F. Bouchet, T. Dauxois and S. Ruffo, J. Stat. Phys. 
\textbf{119}, (2005) 677.

\bibitem{ElskensBook}Y. Elskens and D.F. Escande, \textit{Microscopic
Dynamics of Plasmas and Chaos}, (IoP Publishing, Bristol 2003).

\bibitem{andrea} A. Antoniazzi, G. De Ninno, D. Fanelli, A. Guarino and S. Ruffo,
J. Phys.: Conf. Ser., \textbf{7}, (2005) 143.

\bibitem{Johal} A. Antoniazzi, R. S. Johal, D. Fanelli and S. Ruffo, Comm. Nonlin. Sci. Num. 
Simul., \textbf{13}, (2008) 2.

\bibitem{BarreBouchet} J. Barr\'e and F. Bouchet, J. Stat. Phys. \textbf{118}, (2005) 1073.

\bibitem{CARL} C. von Cube et al., Phys. Rev. Lett., \textbf{93}, (2004) 083601; C. von Cube et al.,
Fortschr. Phys., \textbf{54}, (2006) 726.

\bibitem{Sievers} L.Q. English, M. Sato and A. J. Sievers, Phys. Rev. B, \textbf{67}, (2003) 024403.

\bibitem{Campa} A. Campa, R. Khomeriki, D. Mukamel and S. Ruffo, Phys. Rev. B, \textbf{76}, (2007) 
064415.


\end{thebibliography}
%

\end{document}